\documentclass[12pt]{article}
\usepackage{amssymb,epsfig}

\renewcommand{\thefootnote}{\fnsymbol{footnote}}

\begin{document}

\title{
\begin{flushright}
\ \\*[-80pt] 
\begin{minipage}{0.2\linewidth}
\normalsize
hep-th/0611024 \\
YITP-06-59 \\
TU-780 \\
KUNS-2051 \\*[50pt]
\end{minipage}
\end{flushright}
{\Large \bf 
Moduli stabilization, F-term uplifting and 
soft supersymmetry breaking terms
\\*[20pt]}}

\author{Hiroyuki~Abe$^{1,}$\footnote{
E-mail address: abe@yukawa.kyoto-u.ac.jp}, \ 
Tetsutaro~Higaki$^{2,}$\footnote{
E-mail address: tetsu@tuhep.phys.tohoku.ac.jp}, \ 
Tatsuo~Kobayashi$^{3,}$\footnote{
E-mail address: kobayash@gauge.scphys.kyoto-u.ac.jp} \ and \ 
Yuji~Omura$^{4,}$\footnote{E-mail address: omura@scphys.kyoto-u.ac.jp
}\\*[20pt]
$^1${\it \normalsize 
Yukawa Institute for Theoretical Physics, Kyoto University, 
Kyoto 606-8502, Japan} \\
$^2${\it \normalsize 
Department of Physics, Tohoku University, 
Sendai 980-8578, Japan} \\
$^3${\it \normalsize 
Department of Physics, Kyoto University, 
Kyoto 606-8502, Japan} \\
$^3${\it \normalsize 
Department of Physics, Kyoto University, 
Kyoto 606-8501, Japan} \\*[50pt]}

\date{
\centerline{\small \bf Abstract}
\begin{minipage}{0.9\linewidth}
\medskip 
\medskip 
\small
We study moduli stabilization with F-term uplifting.
As a source of uplifting F-term, we consider spontaneous 
supersymmetry breaking models, 
e.g. the Polonyi model 
and the Intriligator-Seiberg-Shih model.
We analyze potential minima by requiring almost vanishing 
vacuum energy and evaluate the size of modulus F-term.
We also study soft SUSY breaking terms.
In our scenario, the mirage mediation is dominant in gaugino masses.
Scalar masses can be comparable with gaugino masses or 
much heavier, depending on couplings with spontaneous 
supersymmetry breaking sector.
\end{minipage}
}

\begin{titlepage}
\maketitle
\thispagestyle{empty}
\end{titlepage}


\renewcommand{\thefootnote}{\arabic{footnote}}
\setcounter{footnote}{0}

\section{Introduction}

In superstring theory, moduli stabilization is one of important issues 
to study, because their vacuum expectation values (VEVs) determine 
several types of couplings in 4D effective field theory of massless modes 
and through moduli stabilization supersymmetry (SUSY) may break.

Non-perturbative superpotential, e.g. the gaugino condensation, 
is important to stabilize them.
Even if non-perturbative superpotential is generated, 
there is a problem still.
The good candidate for the potential minimum is a supersymmetric 
vacuum, but it leads to a negative vacuum energy if 
the gravitino mass $m_{3/2}$ is non-vanishing.
It is not straightforward to realize the de Sitter vacuum with 
the almost vanishing vacuum energy, where moduli fields are 
stabilized.
One may obtain such vacuum when we consider rather complicated 
superpotential and K\"ahler potential of moduli fields 
including several free parameters.

Recently, in Ref.~\cite{Kachru:2003aw} a new scenario 
was proposed to lead to a de Sitter (or Minkowski) vacuum, 
where all of moduli are stabilized in type IIB string models,
and it is the so-called KKLT scenario.
The KKLT scenario consists of three steps.
At the first step, it is assumed that the dilaton and 
complex structure moduli are stabilized through flux 
compactification \cite{Giddings:2001yu}.
At the second step, non-perturbative superpotential terms, 
which depend on the K\"ahler moduli, are introduced.
Thus, such superpotential stabilizes remaining K\"ahler moduli.
However, that leads to a supersymmetric anti de Sitter (AdS) vacuum.
At the third step, the AdS vacuum is uplifted by 
introducing anti-D3 branes, which break SUSY 
explicitly, at the tip of warp throat.

Phenomenological aspects like soft SUSY breaking terms 
have been studied \cite{Choi:2005ge}.
The KKLT scenario predicts the unique pattern of SUSY breaking 
terms, where modulus mediation and anomaly 
mediation \cite{Randall:1998uk} 
are comparable for gaugino masses, scalar masses and A-terms.
Recently, such type of SUSY breaking mediation is called as 
the mirage mediation, and this pattern of 
s-particle spectrum has significant phenomenological 
implications \cite{Choi:2005uz}-\cite{Choi:2006bh}.

In this paper, we study on the third step of the KKLT scenario, 
i.e. uplifting.
In the KKLT scenario, the uplifting potential is realized by 
explicit SUSY breaking.
Here, we consider the possibility for uplifting 
by spontaneous SUSY breaking within supersymmetric theory, i.e. 
the F-term uplifting.
Such possibility has been studied in Ref.~\cite{Saltman:2004sn}.
Here we study F-term uplifting scenario by use of concrete SUSY
breaking models.
We consider the Polonyi model \cite{Polonyi:1977pj} and 
Intriligator-Seiberg-Shih (ISS) model \cite{Intriligator:2006dd} 
as the source of spontaneous SUSY breaking.
(See also for recent works related with the ISS model 
Ref.~\cite{Franco:2006es}.)
Then, we require almost vanishing vacuum energy.
For this purpose, we fine-tune parameters.
A difference between our approach and the KKLT scenario is that 
in our approach 
the vacuum energy is assumed to vanish within supergravity theory, 
while the vacuum energy is canceled with uplifting energy from 
outside of supergravity in the KKLT scenario.
That might lead to phenomenological aspects different from those in the 
KKLT scenario.
We study soft SUSY breaking terms and show 
the modulus and anomaly mediation are comparable in 
the gaugino masses.
The size of scalar masses depends on how to couple with 
the spontaneous SUSY breaking sector.

This paper is organized as follows.
In the next section, we review on the Polonyi model and 
the ISS model.
We consider the ISS model within the framework of 
supergravity.
In section 3, we combine the moduli stabilization with 
the Polonyi model and the ISS model to realize 
de Sitter (Minkowski) vacuum.
We analyze potential minima and evaluate the magnitude of 
modulus F-term.
In section 4, we study soft SUSY breaking terms.
Section 5 is devoted to conclusion and discussion.

\section{SUSY breaking models}

In this section, we give a brief review on spontaneous 
SUSY breaking models, which shall be used as the source 
of uplifting in the next section.
We consider the Polonyi model and ISS model.
We use the unit with $M_{Pl}=1$, where $M_{Pl}$ denotes the reduced 
Planck scale.

\subsection{Polonyi model}
The Polonyi model is given by the following 
K\"ahler potential $K$ and superpotential $W$: 
\begin{eqnarray}
K &=& |\Phi|^2, \qquad 
W \ = \ c+\mu^2 \Phi, 
\nonumber
\end{eqnarray}
where $\Phi$ is the Polonyi field and $c$ and $\mu^2$ are constants.
The scalar potential 
\begin{eqnarray}
V &=& e^G (G^{I\bar{J}}G_IG_{\bar{J}}-3), 
\qquad 
G \ = \ K+\ln |W|^2, 
\nonumber
\end{eqnarray}
is minimized by a real VEV, 
$\langle \Phi \rangle = \phi$, 
satisfying the stationary condition 
\begin{eqnarray}
V_\Phi \Big|_0 
&=& e^G G_\Phi (G_{\Phi \Phi}+G_\Phi^2-2) \Big|_0 
\ = \ 0, 
\nonumber
\end{eqnarray}
where 
\begin{eqnarray}
(G_{\Phi \Phi}+G_\Phi^2-2) \Big|_0 
&=& \tilde{W}^{-1} (\phi^3+\tilde{c}\, \phi^2-2\tilde{c}), 
\nonumber \\
G_\Phi \Big|_0 &=& \tilde{W}^{-1} (\phi^2+\tilde{c}\,\phi+1), 
\nonumber
\end{eqnarray}
$\tilde{W}=\mu^{-2}W|_0=\phi+\tilde{c}$, 
$\tilde{c}=\mu^{-2}c$, 
and $f|_0$ denotes $f|_{\Phi=\langle \Phi \rangle}$ 
for a function $f=f(\Phi)$. 

If the parameters $c$ and $\mu^2$ are within the range 
$-2<\tilde{c}<2$, we find  $G_\Phi \Big|_0 > 0$ and the 
SUSY point is not a stationary point of the scalar potential. 
Instead, the potential has a SUSY breaking stationary point 
determined by 
\begin{eqnarray}
\phi^3+\tilde{c}\, \phi^2-2\tilde{c} &=& 0. 
\nonumber
\end{eqnarray}
Requiring that this SUSY breaking minimum 
has a vanishing vacuum energy 
\begin{eqnarray}
V \Big|_0 &=& e^G (G_\Phi^2-3) \Big|_0 \ = \ 0, 
\nonumber
\end{eqnarray}
that is 
\begin{eqnarray}
(\phi^2+\tilde{c}\,\phi+1)^2 &=& 3 (\phi+\tilde{c})^2, 
\nonumber
\end{eqnarray}
the VEV $\phi$ and the parameter $\tilde{c}$ are determined as 
\begin{eqnarray}
\phi &=& \sqrt{3}-1, \qquad 
\tilde{c} \ = \ 2-\sqrt{3}. 
\nonumber
\end{eqnarray}
This is the so-called Polonyi solution, which corresponds 
to a SUSY-breaking Minkowski minimum.

\subsection{ISS model in supergravity}

Here, we review briefly on the ISS model and study its supergravity 
extension.
We consider the following K\"ahler and superpotential, 
\begin{eqnarray}
K &=& 
\sum_{i,c}(|\phi^c_{\ i}|^2+|\tilde\phi^i_{\ c}|^2)
+\sum_{i,j}|\Phi^i_{\ j}|^2, 
\nonumber \\
W &=& 
h(\mu^2 \Phi^i_{\ i}-\phi^c_{\ i} \Phi^i_{\ j} \tilde\phi^j_{\ c}), 
\nonumber
\end{eqnarray}
where $i=1,2,\ldots,N_f$,\,$c=1,2,\ldots,N$ and $N=N_f-N_c$. 
This theory is dual to the $SU(N_c)$ theory with $N_f$ flavors of 
``quarks'' $q^i$ and $\bar q_i$, which have the superpotential 
\begin{equation}
W=h\mu^2q \bar q.
\end{equation}

In a global SUSY analysis of Ref.~\cite{Intriligator:2006dd}, 
this model has a SUSY breaking vacuum 
\begin{eqnarray}
\langle \phi \rangle &=& (\phi_0,0), \qquad 
\langle \tilde\phi \rangle \ = \ 
\left( 
\begin{array}{cc}
\tilde\phi_0 \\ 0 
\end{array}
\right), \qquad 
\langle \Phi \rangle \ = \ 
\left( 
\begin{array}{cc}
0 & 0 \\ 0 & \Phi_0 
\end{array} 
\right), 
\nonumber
\end{eqnarray}
where $\tilde\phi_0 \phi_0=\mu^2\,1_N$, and 
$\Phi_0$ is a $(N_f-N) \times (N_f-N)$ matrix. 
The vacuum energy at this minimum is given by 
\begin{eqnarray}
V_0 &=& (N_f-N)|h \mu^2|^2. 
\nonumber
\end{eqnarray} 

Here we study the supergravity extension of this model. 
For simplicity, we consider the case with 
$N_f=2$ and $N_c=1$.
Although the model with $N_c=1$ has no dual theory, 
we can analyze the model with $N_c > 1$ in a similar way.
We parameterize fields as 
\begin{eqnarray}
\phi_i &=& (\chi,\rho), \qquad 
\tilde\phi^i \ = \ 
\left( 
\begin{array}{cc}
\tilde\chi \\ \tilde\rho 
\end{array}
\right), \qquad 
\Phi^i_{\ j} \ = \ 
\left( 
\begin{array}{cc}
Y & z \\ \tilde{z} & X 
\end{array} 
\right). 
\nonumber
\end{eqnarray}
Then the system is described by 
\begin{eqnarray}
K &=& 
|\chi|^2+|\rho|^2+|\tilde\chi|^2+|\tilde\rho|^2
+|Y|^2+|X|^2+|z|^2+|\tilde{z}|^2, 
\nonumber \\
W &=& c+
h \mu^2 (Y+X) 
-h(\chi \tilde\chi Y 
+\rho \tilde\rho X 
+\chi \tilde\rho z
+\rho \tilde\chi \tilde{z}), 
\nonumber
\end{eqnarray}
where we add a constant superpotential term, $c$, 
for generality. 
The minimum obtained in the above global SUSY analysis 
corresponds to 
\begin{eqnarray}
\langle \chi \rangle 
&=& \langle \tilde\chi \rangle 
\ = \ \mu, \qquad 
\langle X \rangle 
\ = \ \Phi_0, 
\nonumber
\end{eqnarray}
and others are all vanishing. 
Note that these VEVs generate a SUSY mass of $O(h \mu)$ 
for most directions such as $\chi+\tilde\chi$ and $Y$, while 
$X$ and the real part of $\chi-\tilde\chi$ receive a SUSY 
breaking mass of $O(h^2 \mu/8 \pi)$ at the 1-loop level. 
In addition, we have some Goldstone modes associated 
with the broken flavor symmetries as shown 
in Ref.~\cite{Intriligator:2006dd}. 
The ISS model corresponds to generalized O'Raifeartaigh model.

On the other hand, supergravity scalar potential is given by 
\begin{eqnarray}
V &=& e^K (K^{I\bar{J}} (D_IW)(D_{\bar{J}}\bar{W})-3|W|^2), 
\nonumber
\end{eqnarray}
where $D_IW=W_I+K_IW$ is given, e.g. as 
\begin{eqnarray}
D_\chi W &=& -h (Y \tilde\chi +z \tilde\rho) +\bar\chi W, 
\nonumber \\
D_\rho W &=& -h (\tilde{z} \tilde\chi + \tilde{\rho}X ) +\bar\rho W, 
\nonumber \\
D_z W &=& -h \chi \tilde\rho + \bar{z} W, 
\nonumber \\
D_Y W &=& -h \chi \tilde\chi  + h \mu^2 +\bar{Y} W. 
\nonumber
\end{eqnarray}
Noticing that conditions 
$D_{\chi,\tilde\chi,\rho,\tilde\rho,z,\tilde{z},Y}W=0$ 
can be satisfied by 
\begin{eqnarray}
\rho,\, \tilde\rho,\, z,\, \tilde{z} &=& 0, \qquad 
hY \ = \ O(W) \ = \ O (m_{3/2}), \qquad 
\chi,\, \tilde\chi \ \simeq \ \mu, 
\nonumber
\end{eqnarray}
we find that, similarly to the global SUSY case, 
most fields are stabilized (prior to $X$) 
by the SUSY condition 
$D_{\chi,\tilde\chi,\rho,\tilde\rho,z,\tilde{z},Y}W=0$ 
due to a large SUSY mass of $O(h \mu)$. 
Although some directions such as pseudo moduli 
and Goldstone modes should remain as light fields in 
the effective potential below, these effects are 
irrelevant to the following analysis of $X$ which 
is essentially the source of SUSY breaking, and 
we will omit the terms involving these modes.

Then, the effective potential for the remaining $X$ 
is described by 
\begin{eqnarray}
V &=& V^{(0)}+V^{(1)}, 
\nonumber \\
V^{(0)} &=& 
e^K (K^{X\bar{X}} (D_X W)(D_{\bar{X}}\bar{W})-3|W|^2), 
\nonumber \\
V^{(1)} &=& m_X^2 |X|^2, 
\nonumber
\end{eqnarray}
where 
\begin{eqnarray}
K &=& |X|^2, \qquad 
W \ = \ c+\mu^2 X. 
\nonumber
\end{eqnarray}
In the superpotential, $c$ and $\mu$ are redefined as 
$c+ \langle (\mu^2 -\chi \tilde\chi) hY \rangle \to c$, 
$\mu$ as $\sqrt{h}\mu \to \mu$, respectively. 
The mass $m_X$ represents a SUSY breaking mass 
for $X$ generated at the 1-loop level. 

The form of the tree level scalar potential $V^{(0)}$ is 
the same as the Polonyi model analyzed in the previous section.
(The Polonyi field $\Phi$ is now replaced by $X$.)
However, we have a SUSY breaking mass for $X$ in $V^{(1)}$. 
We expand the derivative of the scalar potential, 
$V_X=\partial_X V$, under the assumption that 
$c \sim \mu^2$, $\langle X \rangle = x \sim \mu^2$ 
and $\mu^2 \ll 1$ in the unit $M_{Pl}=1$, and find 
\begin{eqnarray}
V_X &=& V^{(0)}_X+V^{(1)}_X 
\ = \ 
\mu^6 e^K W^{-1} \Big( 
\mu^{-4}c\,\big\{ (\mu^{-1} m_X)^2x-2c \big\}
+O(\mu^4) \Big). 
\nonumber
\end{eqnarray}
The stationary condition $V_X=0$ results in 
\begin{eqnarray}
x &\simeq& 2 \left( \frac{\mu}{m_X} \right)^2 c. 
\label{eq:isssugravev}
\end{eqnarray}
Similarly, the scalar potential itself is expanded as 
\begin{eqnarray}
V &=& V^{(0)}+V^{(1)} 
\ = \ 
e^K (m_X^2 x^2-4c \mu^2 x+\mu^4-3c^2+O(\mu^{10})), 
\nonumber
\end{eqnarray}
and the condition for vanishing vacuum energy $V=0$ 
as well as Eq.~(\ref{eq:isssugravev}) leads to 
\begin{eqnarray}
c &\simeq& 
\frac{1}{\sqrt{3}} \mu^2 
\bigg( 1-\frac{2}{3}\, 
\Big( \frac{\mu}{m_X} \Big)^2 \mu^2 \bigg). 
\label{eq:isssugrac}
\end{eqnarray}
These results are consistent with the assumption 
$x,c \sim \mu^2 \ll 1$ by recalling that 
$m_X \approx \mu$. 

The vacuum value of the superpotential evaluated 
by Eqs.~(\ref{eq:isssugravev}) and (\ref{eq:isssugrac}) 
is found to be 
\begin{eqnarray}
W \Big|_0 &=& c+\mu^2 x 
\ = \ \frac{1}{\sqrt{3}} \mu^2 +O(\mu^4), 
\nonumber
\end{eqnarray}
and SUSY is broken at this Minkowski minimum due to 
\begin{eqnarray}
D_X W \Big|_0 &=& (W_X+K_XW) \Big|_0 
\ = \ \mu^2+O(\mu^4). 
\nonumber
\end{eqnarray}

\section{Moduli stabilization and F-term uplifting}

Here, we study stabilization of the modulus $T$ and uplifting.
We assume the other moduli are stabilized at $M_{Pl}$, e.g. 
by flux background, but the single modulus $T$ remains light.
First, we consider the model, where the K\"ahler potential 
and superpotential are obtained as 
\begin{eqnarray}
 & & K= - n_T \ln(T + \bar T), \qquad W=c -A e^{-aT}.
\nonumber
\end{eqnarray}
The second term can be generated by gaugino condensation 
in the hidden sector, and in this case we have $A \sim 1$ and 
$a=O(10)$.
Furthermore, we assume that $c \ll 1$.
Its scalar potential is written as 
\begin{eqnarray}
V = e^K(K^{T \bar T}|D_T W|^2  -3|W|^2).
\nonumber
\end{eqnarray}
The modulus $T$ is stabilized at $D_TW=0$ as 
$a {\rm Re}(T) \simeq \ln A/c$, and the modulus mass is obtained as 
$m_T = am_{3/2}$.
However, we obtain the negative vacuum energy $V=-3e^K|W|^2<0$.
This corresponds to the second step in the KKLT scenario.
In the KKLT scenario, the explicit SUSY breaking effect due to 
anti-D3 brane is added to uplift the potential and to realize 
de Sitter (Minkowski) vacuum.
Here, we do not add such explicit SUSY breaking effect, but 
combine the moduli stabilization and spontaneous SUSY breaking models 
reviewed in the previous section to realize de Sitter (Minkowski)
vacuum.
Then, we study potential minima.

\subsection{Polonyi-KKLT model}
We study a combination of the Polonyi model 
and the KKLT-type model:\footnote{
A similar model has been studied for another reason in 
Ref.~\cite{Dine:2006ii}.}
\begin{eqnarray}
K &=& |\Phi|^2-n_T \ln(T+\bar{T}), \qquad 
W \ = \ c+\mu^2 \Phi-Ae^{-aT}.
\nonumber
\end{eqnarray} 
As in the KKLT model, we assume 
\begin{eqnarray}
A &\sim& 1, \qquad 
a \ \gg \ 1, \qquad 
c,\,\mu^2 \ \ll \ 1, 
\nonumber
\end{eqnarray}
in the unit $M_{Pl}=1$. 

The scalar potential is now a function of two chiral 
superfields $\Phi$ and $T$, and complicated. 
Therefore, we first find a `reference point' which 
seems to be close to the genuine stationary point, and 
then estimate the deviation from the stationary point. 

We define the reference point 
$(\Phi_0, T_0)=(\phi,t)$ 
such that the following conditions are satisfied there:  
\begin{eqnarray}
V_\Phi \Big|_0 &=& 0, \qquad 
(D_\Phi W \Big|_0 \ne 0) 
\nonumber \\
D_T W \Big|_0 &=&  0, 
\nonumber
\end{eqnarray}
where $D_IW=W_I+K_IW$ for $I=(\Phi,T)$, and 
$f|_0=f|_{\Phi=\phi,T=t}$ for a function $f=f(\Phi,T)$. 
The first condition $V_\Phi |_0=0$ is easily solved just 
by interpreting as $\tilde{c}=\mu^{-2}(c-Ae^{-at})$ in 
the previous analysis for the Polonyi model, and we find 
\begin{eqnarray}
\phi &=& \sqrt{3}-1, \qquad 
\mu^{-2}(c-Ae^{-at}) \ = \ 2-\sqrt{3}. 
\nonumber
\end{eqnarray}

The true minimum is represented by 
\begin{eqnarray}
\langle \Phi \rangle &=& \Phi_0+\delta \Phi, \qquad 
\langle T \rangle \ = \ T_0+\delta T, 
\nonumber 
\end{eqnarray}
where $\delta \Phi/\Phi_0 \ll1$ 
and $\delta T/T_0 \ll 1$ are assumed. 
The superpotential and its derivatives are expanded as 
\begin{eqnarray}
W &=& W \Big|_0+\partial_T W \Big|_0 \,\delta T+\cdots 
\ \simeq \ \mu^2+aAe^{-at}\,\delta T, 
\nonumber \\
D_TW &=& D_TW \Big|_0+\partial_T W_T \Big|_0 \,\delta T+\cdots 
\ \simeq \ -a^2Ae^{-at}\,\delta T, 
\nonumber \\
D_\Phi W &=& W_\Phi+K_\Phi W 
\ = \ \mu^2+(\sqrt{3}-1)(\mu^2+aAe^{-at}\,\delta T+\cdots) 
\nonumber \\ &\simeq&
\sqrt{3}\mu^2+(\sqrt{3}-1)aAe^{-at}\,\delta T, 
\nonumber
\end{eqnarray}
where the ellipsis stands for the sub-dominant terms for 
which the coefficients are not enhanced by $a \gg 1$ 
or higher-order terms in powers of $\delta T$. 
From this, we find 
$\bar{F}^{\bar{T}}=-e^{K/2}K^{\bar{T}T}D_TW \simeq 
e^{K/2}K^{\bar{T}T}a^2Ae^{-at}\,\delta T$, and then 
the scalar potential is expanded as 
\begin{eqnarray}
V &=& K_{I\bar{J}}F^I\bar{F}^{\bar{J}}-3e^K |W|^2 
\nonumber \\ &=& 
e^K (|D_\Phi W|^2-3|W|^2)+K_{T\bar{T}}\bar{F}^{\bar{T}}F^T 
\nonumber \\ &=& 
e^K\Big|_0  \Big\{ (D_{\bar{\Phi}}\bar{W})\Big|_0 
(\sqrt{3}\mu^2+(\sqrt{3}-1)aAe^{-at}\,\delta T)
-3 \bar{W} \Big|_0 (\mu^2+aAe^{-at}\,\delta T) \Big\} 
\nonumber \\ &&
+K_{T\bar{T}}e^{K/2}K^{\bar{T}T}\Big|_0 
a^2Ae^{-at}\,\delta T\,F^T 
+\cdots, 
\nonumber
\end{eqnarray}
where $F^I=-e^{K/2}K^{I\bar{J}}D_{\bar{J}}\bar{W}$ 
for $I,J=(\Phi,T)$, and again the ellipsis represents 
sub-dominant terms. 

Noticing $D_\Phi W \Big|_0=\sqrt{3}\mu^2$ and 
$W \Big|_0=\mu^2$, we obtain 
\begin{eqnarray}
V &\simeq& 
e^K (ae^{-K/2}F^T-\sqrt{3}\mu^2)aAe^{-at}\,\delta T. 
\nonumber
\end{eqnarray}
Requiring $\partial V/\partial (\delta T)=0$, the 
F-term $F^T$ is determined as 
\begin{eqnarray}
F^T &\simeq& \sqrt{3}a^{-1} e^{K/2}W \Big|_0, 
\nonumber
\end{eqnarray}
which is suppressed by a large factor $a \gg 1$. 
The order parameter $F^T$ is vanishing at the 
reference point $F^T|_0=0$, and generated by 
the small deviation $\delta \bar{T}(=\delta T)$ as 
$F^T \approx -e^{K/2}K^{T\bar{T}}a^2Ae^{-at}\,\delta \bar{T}$. 
Then, from the above value of $F^T$, we can estimate 
$\delta T$ as 
\begin{eqnarray}
\delta T/T_0 &\approx& 
\frac{1}{(at)^2} \ll 1, 
\nonumber
\end{eqnarray}
where we have adopted 
$W|_0=-K_T^{-1}W_T|_0=-(K_T)^{-1}|_0aAe^{-at}$. 

Then, as expected, the true minimum is very close to the 
reference point. The small deviation $\delta T$ yields 
the small SUSY breaking order parameter $F^T$ compared to 
$F^\Phi$ and $m_{3/2}=e^{K/2}W$. Particularly, the ratio 
between the anomaly mediation and modulus mediation is given by 
\begin{eqnarray}
\alpha &=& 
\frac{m_{3/2}}{\ln (M_{Pl}/m_{3/2})}\,\frac{T+\bar{T}}{F^T} 
\ \simeq \ \frac{at}{2\sqrt{3}\pi^2} 
\ \approx \ \frac{2}{\sqrt{3}}, 
\nonumber
\end{eqnarray}
where $\ln (M_{Pl}/m_{3/2}) \simeq 4\pi^2$ for 
$m_{3/2}=O(10)$ TeV, and 
$at \simeq 4\pi^2$ is determined by $D_TW|_0=0$. 

This is compared to the original KKLT model, 
$\alpha_{KKLT} \simeq 1$. 
We remark that, in the KKLT model, SUSY is broken by the 
anti-D3 brane which generates an explicit SUSY breaking 
in the 4D effective $N=1$ supergravity, while SUSY is broken within 
the $N=1$ supergravity model studied in this section.

\subsection{ISS-KKLT model}

Here, we study a combination of the ISS model 
and the KKLT-type model: 
\begin{eqnarray}
V &=& V^{(0)}+V^{(1)}, 
\nonumber \\
V^{(0)} &=& 
e^K (K^{I\bar{J}} (D_I W)(D_{\bar{J}}\bar{W})-3|W|^2), 
\nonumber \\
V^{(1)} &=& m_X^2 |X|^2, 
\nonumber
\end{eqnarray}
where $I,J=(X,T)$ and 
\begin{eqnarray}
K &=& |X|^2-n_T \ln(T+\bar{T}), \qquad 
W \ = \ c+\mu^2 X-Ae^{-aT}. 
\nonumber
\end{eqnarray}
As in the analysis of Polonyi-KKLT model, we shall find 
a minimum of this model by the perturbation from the 
reference point $(X,T)=(X_0,T_0)=(x,t)$ where 
$V_X|_0=0$, $D_XW|_0 \ne 0$ and $D_TW|_0=0$ are satisfied. 
The first condition $V_X|_0=0$ is solved just by replacing 
$c$ by $\tilde{c}=c-Ae^{-at}$ in the previous analysis of 
the pure ISS model. Then we find 
\begin{eqnarray}
x &=& 2 \left( \frac{\mu}{m_X} \right)^2 \tilde{c}. 
\nonumber
\end{eqnarray}
The true minimum is assumed to be located close to 
the reference point, 
\begin{eqnarray}
\langle X \rangle &=& X_0+\delta X, \qquad 
\langle T \rangle \ = \ T_0+\delta T, 
\nonumber
\end{eqnarray}
where $\delta X/X_0$, $\delta T/T_0 \ll 1$. 

Similarly to the Polonyi-KKLT model, the superpotential 
and its derivatives are expanded as 
\begin{eqnarray}
W &\simeq& 
\frac{1}{\sqrt{3}}\mu^2+aAe^{-at}\,\delta T, 
\nonumber \\
D_TW &\simeq& -a^2Ae^{-at}\,\delta T, 
\nonumber \\
D_XW &\simeq& \mu^2. 
\nonumber
\end{eqnarray}
Then, the scalar potential is given by 
\begin{eqnarray}
V &=& V^{(0)}+V^{(1)} 
\nonumber \\ &\simeq& 
e^K \Big|_0 (m_X^2 x^2-4\tilde{c} \mu^2 x+\mu^4-3\tilde{c}^2)
+aAe^{-at}(ae^{K/2}F^T-\sqrt{3}\mu^2)\,\delta T. 
\nonumber
\end{eqnarray}
The stationary condition $\partial V/\partial (\delta T)$ 
determines $F^T$ as 
\begin{eqnarray}
F^T &\simeq& 3a^{-1}e^{K/2}W \Big|_0, 
\nonumber
\end{eqnarray}
and the condition for vanishing vacuum energy is 
the same as the pure ISS model besides the replacement 
$c \to \tilde{c}$, i.e., 
\begin{eqnarray}
\tilde{c} &\simeq& 
\frac{1}{\sqrt{3}} \mu^2 
\bigg( 1-\frac{2}{3}\, 
\Big( \frac{\mu}{m_X} \Big)^2 \mu^2 \bigg). 
\nonumber
\end{eqnarray}
The anomaly-to-modulus ratio for the SUSY breaking mediation 
in this case is given by 
\begin{eqnarray}
\alpha &\simeq& \frac{at}{6\pi^2} 
\ \approx \ \frac{2}{3}. 
\nonumber
\end{eqnarray}
Note that $\alpha_{ISS-KKLT} \simeq \alpha_{P-KKLT}/\sqrt{3}$.

\subsection{Stringy origin}

In the ISS-KKLT model, the gravitino mass $m_{3/2}$ is determined by 
the constant $c$ and supersymmetric mass $\mu^2$.
To realize low-energy SUSY, we need suppressed values of 
$c$ and $\mu^2$ compared with $M_{Pl}$.
Here we comment on what can be a source of such suppressed terms.
Recall our first assumption, that is, we have assumed that 
all of moduli except the modulus $T$ are stabilized at $M_{Pl}$.
We denote these heavy moduli representatively by $S$.
When these heavy moduli $S$ appear in low-energy effective theory, 
they can be replaced by their VEVs.
It is plausible that non-perturbative effects like gaugino
condensation and string/D-brane instanton effects 
generate 
\begin{eqnarray}
 & & W=Ce^{-\gamma S} + C'e^{-\gamma' S}q \bar q,
\nonumber
\end{eqnarray}
where $C,C'$ are $O(1)$ of constants.\footnote{
The possibility of the first term has been considered 
in Ref.~\cite{Abe:2005rx,Abe:2006xi}, and 
the possibility of the second term has been considered for the Higgs 
$\mu$-term in Ref.~\cite{Choi:2005uz,Choi:2005hd}.
See also Ref.~\cite{Blumenhagen:2006xt}.}
The coefficients $\gamma$ and $\gamma'$ are constants determined by 
discrete numbers, e.g. beta-function coefficients for gaugino 
condensates.
These terms become sources for $c$ and $\mu^2$ when we replace 
$S$ by its VEV.
Thus, when $\gamma \langle S \rangle $ and $\gamma' \langle S \rangle$ 
are of $O(10)$, 
we would have suppressed values of $c$ and $\mu^2$.
Furthermore, if $\gamma = \gamma'$, we expect $c=O(\mu^2)$, 
although we need fine-tuning between $C$ and $C'$ to realize 
almost vanishing vacuum energy.

Furthermore, we have considered the simple case of $T$-dependent 
superpotential, i.e. $W_{np}=Ae^{-at}$ with $A=O(1)$ in the unit 
$M_{Pl}=1$.
However, when the gauge kinetic function of the hidden sector 
is written as a linear combination of $S$ and $T$ and 
this gaugino condensates, we would have the following superpotential term,
\begin{eqnarray}
 & & W_{np}= A e^{-aT-a'S}.
\nonumber
\end{eqnarray}
instead of $W_{np}=Ae^{-aT}$.
Using this superpotential term, we can analyze potential minima 
in a way similar to the previous section.
Then, we have various values of $\alpha$ depending on a value of $a'S$
like Ref.~\cite{Abe:2005rx}, but its order would be obtained as 
$\alpha =O(1)$.

\section{Soft SUSY breaking terms}

Here we study soft SUSY breaking terms of the visible sector.
The F-term of modulus $F^T$ is smaller than the gravitino mass 
$m_{3/2}$ by $O(1/8\pi^2)$.
That is, the modulus mediation and anomaly mediation are 
comparable, and its ratio $\alpha$ is of $O(1)$.

First, we evaluate gaugino masses, whose gauge 
kinetic functions are obtained as 
\begin{eqnarray}
 & & f_a = k_a T + \Delta f_a,
\nonumber
\end{eqnarray}
where $k_a$ and $\Delta f_a$ are constants and $\Delta f_a$ may depend on 
heavy moduli.
The $F^T$ contribution to gaugino mass is obtained as 
\begin{eqnarray}
 & & M_a ^{(T)}= \frac{k_a}{f_a + \bar f_a}F^T.
\nonumber
\end{eqnarray}
In addition, there is the contribution from anomaly mediation,
\begin{eqnarray}
 & & M_a^{(AM)}= -\frac{\beta_{g^2_a}}{2g^2_a}m_{3/2},
\nonumber
\end{eqnarray}
where $g_a$ is the gauge coupling and 
$\beta_{g^2_a}$ is the beta-function of $g^2_a$.
When $\alpha=O(1)$, these two contributions are comparable.
That is the mirage mediation.
If the Polonyi field and the field $X$ in the ISS model appear 
in the gauge kinetic function, the situation would change.

Similarly, we evaluate scalar masses $m_i$ of chiral multiplets $Q^i$ 
in the visible sector.
We may have several types of possibilities for assuming 
K\"ahler metric of chiral multiplets  $Q^i$, in particular, how 
$Q^i$ couple with the Polonyi field $\Phi$ and the field $X$ in 
the ISS model.
Here we consider three models.
In the model I, the visible matter is separated from $\Phi$ and $X$ in 
the form of $Y_i = e^{-K/3}Z_{i \bar i}|Q^i|^2$, 
where $Z_{i \bar i}$ is the K\"ahler metric of visible fields 
$Q^i$.\footnote{ The authors would like to thank K.~Choi for pointing 
this possibility.}
Such assumption could be realized by the setup that 
the SUSY breaking source is localized far away from the visible 
matter fields in the compact space.
In the model II, the  K\"ahler metric $Z_{i \bar i}$  depends on only
$T$ and $\bar T$, but not $\Phi$ or $X$.
Thus, the visible modes $Q_i$ are not sequestered from $\Phi$ and $X$ in 
$Y_i$.
In the model III, the K\"ahler metric $Z_{i \bar i}$ includes 
a contact term like $c_i|X|^2|Q^i|^2/M_p^2$. 
The visible matter fields $Q_i$ are not sequestered from $\Phi$ or 
$X$ in $Y_i$ or $Z_{i \bar i}$.
We assume the almost vanishing vacuum energy, $V_0 \simeq 0 $, 
in evaluation of scalar masses for all of three models.


In the model I, soft scalar masses squared are obtained at the
tree-level as 
\begin{eqnarray}
 & & m_i^2 = - |F^T|^2\partial_T \partial_{\bar T} 
\ln Y_{i \bar i},
\end{eqnarray}
where $Y_{i \bar i} = e^{-K/3}Z_{i \bar i}$.
The F-term $F^T$ is suppressed compared with $m_{3/2}$.
Thus, the anomaly mediation is comparable with this tree-level effect.
Then, we have the mirage mediation in soft scalar masses, too.

In the model II, soft scalar masses squared are obtained 
\begin{eqnarray}
 & & m_i^2 = m_{3/2}^2 - |F^T|^2\partial_T \partial_{\bar T} 
\ln Z_{i \bar i}.
\nonumber
\end{eqnarray}
Since the F-term $F^T$ is suppressed compared with $m_{3/2}$, 
the first term is dominant in this case.
Although there is the contribution from anomaly mediation, 
it is sub-dominant.
Thus, in the model II, soft scalar masses are universal, 
\begin{eqnarray}
 & & m_i^2 = m_{3/2}^2.
\nonumber
\end{eqnarray}
In the model III,  there is a contact term like $c_i |X|^2|Q^i|^2/M_p^2$, 
soft scalar masses $m_i$ also depend on $c_i |F^X|^2$.
At any rate, the order of $m_i$ is of $O(m_{3/2})$. 
In both models II and III, scalar masses are quite heavy compared 
with gaugino masses, and these would have radiative corrections.

Moreover, trilinear couplings of scalar fields, i.e. the so-called
A-terms, can be calculated as 
\begin{eqnarray}
 & & h_{ijk} = \sum_I F^Ie^{K/2}[(\partial_I + K_I)Y_{ijk} - 
Y_{ijk}\partial_I\ln(Z_{i \bar i} Z_{j \bar j} Z_{k \bar k}) ],
\nonumber
\end{eqnarray}
where $Y_{ijk}$ is the corresponding Yukawa coupling, and the index $I$ 
includes the modulus $T$ and the Polonyi field $\Phi$ and the field
$X$ of the ISS model.
In the ISS-KKLT model, the VEV of $X$ is suppressed and $K_X$ is suppressed.
Thus, the natural order of $h_{ijk}/Y_{ijk}$ 
would be of $O(F^T/T)$.
In this case, the anomaly mediation is also important and 
the size of $h_{ijk}/Y_{ijk}$  is comparable with the gaugino masses.
On the other hand, if the Yukawa coupling $Y_{ijk}$ includes 
$T$-modulus like 
\begin{eqnarray}
 & & Y_{ijk} \sim e^{-a'_{ijk}T} .
\nonumber
\end{eqnarray}
the contribution from $F^T$ can be enhanced by the term 
$\partial_T Y_{ijk}$, and its order would be of $O(m_{3/2})$.
In the model that all of Yukawa couplings are given in the above form, 
large values of $|a'_{ijk}|$ correspond to suppressed Yukawa couplings.
Thus, in such model, small Yukawa couplings like the first and second
families would correspond to large A-terms of $O(m_{3/2})$, 
while large Yukawa couplings like the third family 
would correspond to smaller A-terms.

In the Polonyi-KKLT model, the VEV of Polonyi field is of $O(M_{Pl})$,
and the dominant contribution to 
$h_{ijk}/Y_{ijk}$ would be obtained as 
\begin{eqnarray}
 & & \frac{h_{ijk}}{Y_{ijk}} = F^\Phi e^{K/2}K_\Phi + F^Te^{K/2}
\partial_T \ln Y_{ijk}.
\end{eqnarray}
Then, it is naturally of $O(m_{3/2})$.
However, when the Polonyi field $\Phi$ is 
sequestered from $Q_i$ in $Y_i$, i.e. 
the model I, there is no contribution from $F^\Phi$.
Thus, when $\partial_T \ln Y_{ijk} =O(1)$, 
the size of $h_{ijk}/Y_{ijk}$ would be comparable with 
the gaugino masses even in the Polonyi-KKLT model.

The magnitudes of the Higgs $\mu$-term and the so-called $B$-term 
depend on how to generate the $\mu$-term.
Suppose that the $\mu$-term is generated as 
\begin{eqnarray}
 & & W_\mu = C e^{-hT}H_u H_d.
\nonumber 
\end{eqnarray}
In this case, the would-be dominant part of $B$-term is obtained, 
e.g. in the ISS-KKLT model as 
\begin{eqnarray}
 & & B=m_{3/2} - h F^T.
\nonumber
\end{eqnarray}
Thus, the natural scale of $B$ is of $O(m_{3/2})$.
If two terms $m_{3/2}$ and $hF^T$ with $h=O(10)$ are canceled each other, 
then $B$ can be of $O(m_{3/2}/8\pi^2)$.
Indeed such cancellation happen in a certain case 
\cite{Choi:2005hd}.

As a result, the full s-particle spectrum of our model is as follows.
We choose the over-all mass scale such that the gaugino masses are 
of $O(100)-O(1000)$ GeV.
When visible matter fields are sequestered from the spontaneous 
SUSY breaking sector, sfermion masses are of the same order as 
gaugino masses, and the gravitino are of $O(10)$ TeV and the mass of 
modulus $T$ is of  $O(100)$ TeV.
When visible matter fields are not sequestered from the spontaneous 
SUSY breaking sector, the gravitino and sfermion masses are of $O(10)$ TeV.
The size of A-terms can be of the same order as gaugino masses or 
gravitino masses, depending on $T$-dependence of Yukawa couplings and 
the sponataneous SUSY breaking mechanism.
The natural scale of $B$-term is of $O(m_{3/2})$, but 
in a certain case we could obtain smaller value of $B$.

\section{Conclusion and discussion}

We have studied modulus stabilization with F-term uplifting.
As explicit models, we have used the Polonyi model and 
the ISS model.
Combining these spontaneous SUSY breaking models and 
the KKLT type of superpotential, we have analyzed 
potential minima.
At the potential minima, the size of modulus F-term $F^T$ is similar to 
the KKLT model, and suppressed by a factor of $O(10)$ 
compared with the gravitino mass.
We have also studied soft SUSY breaking terms.
In the gaugino masses, modulus mediation and anomaly mediation 
are comparable, i.e. the mirage mediation.
On the other hand, sfermion masses can be of the same order as 
the gaugino masses or of $O(m_{3/2})$, depending couplings with 
the spontaneous SUSY breaking sector.
Thus, in the low-energy SUSY scenario, the gaugino masses are 
of $O(100)-O(1000)$ GeV, while sfermion masses are of the same order 
or $O(10)$ TeV. 
The gravitino mass is of $O(10)$ TeV and the modulus $T$ is much heavier.
The A-terms can be of the same order as the gaugino masses 
or gravitino masses.
We have studied two explicit models with F-term uplifting, 
but these spectra would be expected of generic feature of 
F-term uplifting scenario.
The natural size of $B$-term would be of $O(m_{3/2})$, but 
in a certain case it could be much smaller.

Recently, phenomenological aspects of the mirage mediation 
have been studied.
However, spectra of our models, in particular 
sfermion masses as well as A-terms, can differ from 
the mirage mediation.
It would be interesting to study phenomenological aspects of 
models, where the mirage mediation is dominant for 
the gaugino masses and sfermion masses are much heavier.\footnote{
See e.g. Ref.~\cite{Wells:2004di,Ibe:2006de}, where phenomenological 
aspects of similar s-particle spectra have been studied.}

\vskip 0.5cm
{\large \bf Note to be added}

While this paper was being finished, Ref.~\cite{Dudas:2006gr}
appeared, where they also studied the same model as one of ours.

\subsection*{Acknowledgement}
The authors would like to thank F.~Brummer, K.~Choi, 
T.~Goto, A.~Hebecker, K.~Izawa, T.~Onogi and M.~Trapletti
for useful discussions.
H.~A.\/,  T.~H.\/ and T.~K.\/ are supported in part by the
Grand-in-Aid for Scientific Research \#182496, \#171643 
and  \#17540251, respectively.
T.~K.\/ is also supported in part by 
the Grant-in-Aid for
the 21st Century COE ``The Center for Diversity and
Universality in Physics'' from the Ministry of Education, Culture,
Sports, Science and Technology of Japan.

\end{document}